**Title:** The genomic consequences of hybridization

**Authors:** Benjamin M Moran[1,2*+], Cheyenne Payne[1,2*+], Quinn Langdon[1], Daniel L Powell[1,2], Yaniv Brandvain[3], Molly Schumer[1,2,4+]

**Affiliations:**
[1]Department of Biology, Stanford University, Stanford, CA, USA
[2]Centro de Investigaciones Científicas de las Huastecas "Aguazarca", A.C., Calnali, Hidalgo, Mexico
[3]Department of Ecology, Evolution & Behavior and Plant and Microbial Biology, University of Minnesota, St. Paul, MN, USA
[4]Hanna H. Gray Fellow, Howard Hughes Medical Institute, Stanford, CA, USA

*Contributed equally to this work
+Correspondence: benmoran@stanford.edu, cypayne@stanford.edu, schumer@stanford.edu

**Abstract**

In the past decade, advances in genome sequencing have allowed researchers to uncover the history of hybridization in diverse groups of species, including our own. Although the field has made impressive progress in documenting the extent of natural hybridization, both historical and recent, there are still many unanswered questions about its genetic and evolutionary consequences. Recent work has suggested that the outcomes of hybridization in the genome may be in part predictable, but many open questions about the nature of selection on hybrids and the biological variables that shape such selection have hampered progress in this area. We discuss what is known about the mechanisms that drive changes in ancestry in the genome after hybridization, highlight major unresolved questions, and discuss their implications for the predictability of genome evolution after hybridization.

**Introduction**

Recent evidence has shown that hybridization between species is common. Hybridization is widespread across the tree of life, spanning both ancient and recent timescales and a broad range of divergence levels between taxa [1–10]. This appreciation of the prevalence of hybridization has renewed interest among researchers in understanding its consequences.

Perhaps one of the most surprising outcomes of this recent research is the extent to which hybridization shapes the genomes of extant **species** (see **Glossary**). In humans, ~2–5% of the genomes of some populations are derived from ancient **admixture** with our extinct relatives, the Neanderthals and Denisovans [11], including genes that contribute to adaptation and genetic diseases [12–15]. In other taxa, such as swordtail fishes, Italian sparrows, sunflowers, and cichlid fishes, upwards of 10% of some species' genomes are derived from ancient hybridization [16–18]. These findings have spurred interest in the genomic consequences of hybridization.

Some of this genetic exchange reflects the process of **adaptive introgression**, which has been well-documented in several species [19–21]. However, the introgression of adaptive and neutral variants occurs against the backdrop of broad, genome-wide selection against hybrids [22–25] and hybridization-derived regions in the genome [3,26–29]. The mechanisms resulting in lower fitness of hybrids are diverse, ranging from **ecological selection** against hybrids, to differences in the number of deleterious variants harbored by the hybridizing species (known as **hybridization load**), to negative interactions between genes derived from the two parental species' genomes (**hybrid incompatibilities**). Superficially, widespread selection against foreign ancestry seems to conflict with evidence that hybridization is common. However, understanding the processes through which genomes resist ongoing introgression (in the case of **hybrid zones** or **tension zones**) or stabilize after pulses of hybridization can help us reconcile both observations.

Because many factors interact simultaneously in hybrids, genome evolution is unusually dynamic after hybridization. In the last several years, the community has shifted from describing the presence of admixture in the genomes of diverse species to documenting patterns of local variation in ancestry along the genome [3,11,28–33]. One common observation from these cases is that on average selection acts to remove ancestry from the **minor parent** (i.e., the species from which hybrids derive less of their genome) in the most functionally important regions of the genome. However, we still lack a basic understanding of the different forces driving variation in local ancestry, how they interact, and how predictable the ultimate outcomes of hybridization are.

Here, we synthesize the emerging "principles" of hybridization – that is, repeated outcomes observed across species – and outline outstanding questions. In doing so, we focus on the many cases where hybridization appears to globally reduce fitness, even if adaptive introgression occurs locally in the genome [11,34,35], rather than cases where hybridization appears to fuel diversification as has been reported in some systems [17,19,36,37]. We also focus our discussion throughout the manuscript explicitly on ancestry variation across the genome [38,39] rather than on statistics summarizing population differentiation that can be correlated with ancestry [40].

One major challenge for researchers studying the genetic consequences of hybridization is reconciling how different genetic and evolutionary processes may interact in hybrids to shape variation in ancestry along the genome. Most current models consider sources of selection in isolation, but in nature, multiple selective and demographic processes operate simultaneously, potentially interfering with or amplifying each other. We propose that a key priority for future work should be developing predictions about how particular combinations of selective pressures will impact local ancestry patterns after hybridization.

With better models for how selection operates in admixed genomes, we can begin to ask whether outcomes of hybridization between species are in part predictable and where we expect these predictions to break down. In addition to leading to a clearer understanding of the architecture of modern genomes, pursuing these questions will allow us to move from describing patterns of local ancestry variation along the genome to pinpointing the evolutionary and genetic processes driving this variation.

**Models of hybridization and introgression**

In this review we synthesize research of the consequences of admixture spanning different timescales and population histories of hybridization. The advent of inexpensive whole genome sequencing has allowed for the detection of ancient hybridization events, adding to a rich literature of contemporary hybridization events [41–47] and stable zones where hybridization has been ongoing for thousands of generations [6,48–55]. The distinct timescales of different admixture events allow us to ask questions about both the early and late stages of genome evolution and stabilization following hybridization.

Much of our discussion focuses on a pulse model of hybridization, where an admixed population arose from two diverged populations at some time in the past, and has since stopped.

While this model is an oversimplification in most cases, many scenarios of introgression can be well approximated by a pulse model [41–45], which simplifies interpretation of the dynamics of genome stabilization following hybridization. For example, this model lends itself to evaluating how the genome stabilizes over time, which evolutionary processes occur shortly after initial gene exchange [28,56], and which occur over a longer time period [4,11,44,57,58].

While pulses of admixture are common, they are not the only mode of admixture. A rich theoretical tradition has examined ongoing gene flow between two distinct populations (a two-island model) and in spatially structured populations (ecotones and tension zones), and there are many empirical examples that are well approximated by these models [6,48,49,59–61]. Often the intuition and results developed from the pulse model can be effectively brought to bear on these more complex models. For example, some of the processes acting on recent migrants in tension zones approximate those occurring shortly after a pulse of admixture [62,63]. In other cases, tension zones provide a complementary view of the architecture of selection on hybrids, such as which regions of the genome are intolerant to introgression [64–68]. Throughout our discussion below, we reference both literatures, and highlight cases where results from a pulse model can and cannot be extended to island and spatial models of ongoing introgression.

**Emerging principles of hybridization**

Why do some regions of the genome retain genetic material derived from hybridization while others are purged of foreign DNA? We begin here by outlining emerging principles associated with variation in ancestry in admixed genomes, regardless of the evolutionary process driving this variation (see next section). We note that these principles apply to the large number of cases in which selection on average acts against hybridization, but may not apply to systems where hybridization is globally neutral or beneficial [17,36,37,69].

*Principle 1: A combination of rapid and slower removal of foreign ancestry stabilizes admixed genomes*

Variance in genome-wide ancestry in admixed populations is highest just after hybridization and decreases over time as recombination breaks down long ancestry tracts. When foreign ancestry is deleterious, selection during this initial period rapidly reduces the population's admixture proportion [27,70]. This initial "fast" period of purging lasts tens of generations [70], shifts ancestry genome-wide [28,56], and begins to generate broad scale differences in ancestry within and among chromosomes. Populations then enter a "slow" period of purging, where selection on individual hybridization-derived **haplotypes** only subtly shifts genome-wide ancestry proportions. The shape and rate of this change in ancestry can vary from species to species [70], primarily as a function of the total recombination rate (see *Principle 3*). These predictions have been explored primarily under models of pulses of admixture but similar patterns are expected for recent migrants in hybrid zone models [62,63].

*Principle 2: Functionally important regions of the genome experience different rates of introgression*

Although the sources of selection on hybrids undoubtedly differ between species [26,28,31], studies across diverse taxa have largely found that regions of the genome that are dense in coding or conserved elements tend to be particularly resistant to movement between species [3,6,7,29,31,71–73]. In the case of conserved regulatory elements in humans, this pattern is stronger at enhancers that harbor derived mutations as opposed to ancestral variants [74]. The consistency of the observation that introgression is depleted in functionally important regions implies that selection against minor parent ancestry generates barriers to introgression that are, in many cases, common, functionally broad, and **polygenic** [75]. These genome-scale observations echo classic work reporting depleted introgression on sex-chromosomes [66,71,76] and asymmetry in the effects of hybridization between the sexes (e.g. [77–79]; Haldane's rule), well-accepted rules in the speciation literature [72] that have been thoroughly reviewed elsewhere [76,80].

*Principle 3: The recombination landscape plays a key role in genome stabilization*

Selection acts to remove many introgressed haplotypes after hybridization. Because haplotypes derived from the minor parent species are longer in regions of the genome where recombination events are rare, minor parent haplotypes in low recombination rate regions are more likely to harbor variants that will be harmful in hybrids. This is conceptually similar to the reason why ancestry proportions shift drastically in the early generations after hybridization when ancestry tracts are long (i.e. *Principle 1*). Even after genome-wide admixture proportions have stabilized, theory predicts that minor-parent ancestry will be more fully removed from regions of the genome with low recombination rates [81,82]. Data from diverse species, including swordtail fish, humans, monkeyflowers, maize, and *Heliconius* butterflies [28–31,83], support this theoretical prediction (but see [84]). However, differing correlations between recombination rate and gene density can lead to local differences in minor parent ancestry, depending on where in the genome recombination primarily occurs. For example, in humans, recombination rates tend to be locally reduced near genes [85,86], resulting in a tendency to purge introgressed DNA near genes driven by *both* Principles 2 & 3, while in swordtail fish and birds, recombination rates are elevated near genes [87,88], pitting these rules against one another. In fact, the rapid evolution of the recombination landscape [89,90] may be another factor contributing to variation in the landscape of introgression across species groups.

**From pattern to process: Genome evolution after hybridization is shaped by diverse evolutionary forces**

Admixed genomes are a mosaic of regions with little to no minor parent ancestry and regions where such ancestry is much more common. The observed ancestry variation in these modern genomes is likely driven in part by each of the principles described above, which are expected to act whenever there is global selection against hybrid ancestry. The next key question is what demographic processes and mechanisms of selection have generated the rugged ancestry landscape we observe many generations after initial hybridization? We are now poised to address this question, which has been at the heart of research in evolutionary genetics for decades [91], by leveraging data from both ancient and recent hybridization events across diverse groups of species.

Because hybridization combines two diverged genomes into a single organism, hybrids can face a host of challenges, from reconciling protein interactions at the cellular level [92,93] to targeting the appropriate ecological niche at the organismal level [23]. Although we know that reconciling these challenges often involves changes in ancestry at genes and regulatory regions (*Principle 2*), we rarely know the mechanisms that act to drive these changes. Historically, researchers have focused on the possible role of hybrid incompatibilities as a major cause of reduced fitness in hybrids. However, recent work has revealed that other forms of selection, such as hybridization load, can generate similar patterns in hybrid genomes [94]. Determining what different patterns of ancestry can tell us about the sources of selection acting after hybridization is a key challenge for this field.

Although disentangling the causes of selection against introgression is a major goal of the field (**Box 1**) and motivator for our work, we caution readers against drawing a bright line separating some of the models discussed below. This is particularly true for Fisher's Geometric model (see below), which was proposed as a synthetic framework to interpret and predict many patterns and processes underlying hybrid fitness. As such, we approach these models as a source of biological inspiration for the types of mechanisms shaping hybrid genome evolution.

*Hybrid Incompatibilities*

Dobzhansky-Muller hybrid incompatibilities (**DMIs**) occur when mutations that have arisen in each parental species' genome interact negatively in hybrids. DMIs are the best documented and best understood mechanism of selection on hybrids. Indeed, the search for DMIs predates the recognition of the ubiquity of hybridization [95]. In addition to incompatible substitutions that arise in directly interacting proteins, DMIs can also take the form of reciprocal losses following gene duplication or modifications in co-evolving regulatory elements, among other mechanisms [96–99]. The DMI model is conceptually similar to models of 'developmental systems drift', where neutral changes in protein-protein interaction networks can lead to molecular pathways that are incompatible without changing the function of the pathway in either species [100–102]. The loci involved in DMIs identified to date are functionally diverse [103–105], but existing theory and data have hinted at broader evolutionary forces that drive the emergence of hybrid incompatibilities.

DMIs are largely expected to locally restrict gene flow by preventing introgression at the incompatible loci and regions linked to them [6,106,107,107,108], but can also favor the adaptive introgression of pairs of compatible alleles [109]. Thus, the genomic location of DMIs and the forces that drive their evolution will directly impact where in the genome introgression can occur. One well-established example is the observation that introgression is reduced on the sex chromosomes, presumably because DMIs are overrepresented, due to factors such as faster X evolution, meiotic drive, and the importance of X chromosome genes in male fertility [58,76,104,110,111]. Beyond sex chromosomes, certain genes appear to be repeatedly involved in hybrid incompatibilities (Fig. 1; [58,105,112–117]). While some of this overrepresentation may reflect sampling biases [113], as DMIs are characterized across more species it will become increasingly possible to test the hypothesis that certain genes act as "hotspots" for the formation of hybrid incompatibilities. Looking forward, unanswered questions about the number of DMIs that distinguish recently diverged species, the strength of selection acting on them [118], and the rate at which they evolve [119] will be crucial in distinguishing signatures of selection against DMIs from other forms of selection on hybrids.

*Hybridization load*

Historically researchers have considered selection against introgression to reflect interactions between diverged genomes. However, processes occurring within populations can also generate barriers (or thoroughfares) to introgression [26,27,120]. In other words, selection on introgressed ancestry might reflect the unconditional deleterious effect of a mutation, rather than its poor interaction with other sites in the genome (as seen in DMIs). Such mildly deleterious alleles will preferentially reach fixation in populations with weaker purifying selection, such as those with smaller effective population sizes. With sufficient time, a large number of weakly deleterious mutations can accumulate within a species, generating a strong selective force after hybridization with a species that harbors fewer such mutations. Although each mutation is weakly selected against individually, in aggregate these mutations strongly reduce hybrid fitness relative to populations with fewer deleterious mutations, because they are linked to the same haplotypes after hybridization. Interestingly, this prediction holds even if the census population size of the admixed population is relatively small [28]. In the case of a pulse model of admixture, after genome-wide admixture proportions have equilibrated, selection against specific deleterious sites may still drive long-term ancestry purging (see *Effects of hybrid demography*). Empirical studies support this prediction, showing that ancestry from the species with less effective purifying selection can be depleted over many generations of selection, particularly in coding and conserved non-coding regions of the genome [26,27].

In contrast to other models, which predict widespread selection against minor parent ancestry, the additive hybridization load model predicts selection for directional introgression from the species that harbors fewer deleterious mutations. Alternatively, if deleterious mutations are recessive, theory and some empirical data predict that selection will favor an excess of foreign ancestry (**Box 2**; [27,121]), especially in the species with lower historical population sizes. In principle, selection against hybridization load could produce patterns that are distinguishable from other models of selection on hybrids because these weakly deleterious mutations are expected to be broadly distributed throughout the genome and fall within a particular range of selection coefficients (**Box 1**; [26,27]).

*Ecological selection*

Ecological selection is a potentially important but poorly understood source of selection on hybrids. This is in part because less is known about both the **genetic architecture** of ecological adaptation and the ways in which ecological traits can become decoupled in hybrids. Moreover, this source of selection is sensitive to the environments in which hybrids find themselves.

Hybrids may express ecological traits that are intermediate to those of their parent populations (e.g. [18,122]) or express "phenotypically mismatched" traits ([123] and **Box 2**). In such cases, ecological selection will disfavor hybrids [124–126], unless hybrids exist in an intermediate ecological niche or an environment favoring these mismatched phenotypes [127–130]. Like the DMI and hybridization load models, ecological selection is predicted to result in biased ancestry around functionally relevant genomic regions, though the expected direction of bias depends on the environment (**Box 1**).

What patterns of ancestry can indicate the presence of ecological selection on hybrids? The answer to this question depends largely on the architecture of ecological traits [131–134]. While it is straightforward to make predictions about the outcome of ecological selection in hybrids when the trait in question is controlled by a handful of genes, we know less about ancestry shifts after hybridization in ecologically relevant traits with a highly polygenic basis. Theory has explored how traits with a polygenic genetic architecture respond to different types of selection within a species [135,136], but these models do not capture the increased trait variance and **ancestry linkage disequilibrium** expected in hybrids (see next section).

Our discussion of ecological selection on hybrids above ignores "transgressive" segregation – where hybrid trait values fall outside of the range of phenotypes observed in either parent [129]. We discuss the possible interaction of ecological selection and transgressive segregation in **Box 2**.

*Polygenic selection on hybrids*

Given that populations evolve independently before admixture, hybridization has the potential to decouple suites of co-adapted alleles originally linked within the parental species. In hybrids, selection on polygenic traits has been frequently modeled using **Fisher's geometric model** [137], a simple mathematical description of the distance of an individual from its phenotypic optimum, that predicts many of the dynamics of selection against hybrids [94,100–102,138–141]. We note that because of its generality, Fisher's geometric model has also been used to model selection on DMIs among other phenomena, but focus on its application to polygenic traits here.

In a Fisherian model of polygenic adaptation, individual fitness in the parental species can be described as a function of distance from a phenotypic optimum in quantitative trait space, and isolated populations maintain their respective optima through the independent fixation of sets of trait-increasing and trait-decreasing alleles [138]. Crucially, given enough time, the sets of loci underlying the trait and the sign of their phenotypic effects are likely to differ across populations, even between populations with identical phenotypic optima [142,143]. In hybrids, recombination decouples these sets of parental alleles. This can result in hybrid phenotypes that fall outside of the phenotypic optima of either parental species, reducing fitness through a phenomenon known as **segregation load** [144,145]. More precisely, when parental alleles are mixed into different genetic backgrounds, hybrids can show greater variance in a trait than observed in either of the parental species (Fig. 2). If the trait is also under stabilizing selection in hybrids, this increased variance could drive purging of minor parent ancestry over time. Notably, these predictions should hold when parental species are adapting to similar [146] or distinct [147] phenotypic optima (Fig. 2), and when genotypic effects are non-additive [139,148].

*Adaptive introgression against the genomic background*

There is no doubt that haplotypes introduced by hybridization can confer an adaptive advantage [20,21,34,35,149]. Given broad selection against minor parent ancestry, however, adaptive introgression is often occurring against a background of genome-wide purging [29,30,150]. In some cases, this has made adaptively introgressed haplotypes easy to identify empirically since they form peaks of high minor parent ancestry against a background of low minor parent ancestry [3,151]. This also means that several factors will impact the probability that a globally adaptive allele will introgress between species. These include the locations of potentially adaptive alleles relative to deleterious neighbors, the relative selection coefficients on adaptive and deleterious sites, and other features of genome organization [70]. Although there is little empirical or theoretical work in this area to date, some predictions can be made from first principles. For example, an adaptive haplotype in a region of the genome with a very low recombination rate would have a lower probability of introgressing than a haplotype with the same advantage in a higher recombination rate region (*Principle 3*).

These factors highlight that beyond distinguishing between sources of selection on hybrids (**Box 1**), another difficult hurdle is characterizing how they may interact. Although research to date has largely focused on each mechanism in isolation, most hybridization events likely involve the interplay between several modes of selection. As discussed above, in the admixture event between humans and Neanderthals, both hybridization load and adaptive introgression have shaped Neanderthal ancestry in modern human genomes [3,26,27]. This combination of positive and negative selection on hybridization-derived haplotypes can generate interference, especially in the early generations following hybridization when long haplotypes of each ancestry type are common (Fig. 3). Simulations hint that it may be possible to disentangle different signals of selection on hybrids using local ancestry variation (Fig. 3) or changes in ancestry over time (Fig. 4) [152], presenting exciting opportunities for future work.

*Overlooked complexities of selection on hybrids*

The mechanisms discussed above likely represent an incomplete picture of the breadth of forms of selection on hybrids. For example, weak but pervasive epistatic interactions (e.g. of interacting genes in pathways) could select for similar shifts in ancestry as expected from selection on polygenic traits, but whether such weak epistatic interactions are common is unknown. There are also emerging examples of hybrid dysfunction that do not fit cleanly into the sources of selection described above. Such cases raise the question of whether these mechanisms are truly distinct or more often represent a combination of selective forces.

For example, in the case of hybrid gene regulation two frameworks of selection may be simultaneously applied to the same genes. Often under stabilizing selection within the parental species, it is common for *cis*- and *trans*-acting regulatory factors to show evidence of compensatory evolution within species [96,153]. As a result, mismatches in these interacting factors in hybrids can lead to dramatic under- or overexpression of the genes they regulate (Fig. 5). We speculate that this type of misexpression could result in two forms of selection on hybrids. Large-effect expression aberrations would be selected against as a DMI, via selection acting against heterospecific allelic combinations at *cis*- or *trans*-acting loci. For example, allelic combinations that reduce or eliminate expression of a given gene (Fig. 5) can lead to strong selection on this non-functional genotype combination. After the misexpression is resolved, additional smaller effect variants from the two parental species may still have an impact on variance in expression (e.g. Fig. 2B). Changes in ancestry at these variants would then be driven by stabilizing selection on the overall expression of the gene.

Such "priority" effects of selection on hybrids, with rapid purging of interactions in response to strong selective pressures and slower purging associated with weaker selective pressures are reminiscent of the fast versus slow purging of ancestry tracts after initial hybridization (*Principle 1*). While in many cases there is no bright line between the mechanisms of selection discussed in this section, we propose that this approach of considering phases of selection on hybrids may be a fruitful way of understanding the complexity of several intertwined selective forces acting on hybrid genomes.

Because of this complexity, it is important to also note that to some extent the distinctions made between different sources of selection can be arbitrary and not biologically meaningful. Some types of selection on hybrids may be best described by multiple nested mechanisms, as discussed above, whereas others may be innately coupled – such as a DMI that caused by genes underlying an ecologically relevant trait [23,123].

**Predicting the landscape of introgression within and between species**

In the previous sections we discussed what is known about the outcomes of hybridization across diverse species (*Principles 1-3*) as well as the challenges and prospects for understanding how different evolutionary processes lead to changes in ancestry after hybridization. Armed with these tools, we can begin to explore the directions that these advances will allow geneticists and evolutionary biologists to pursue.

*Causes of convergent patterns of introgression across taxa*

Biologists have long been fascinated with the question of whether evolution is predictable [154]. A key unanswered question is the extent to which we can predict outcomes of hybridization within and between pairs of species. At a broad scale, some predictions can be made due to the interplay between selection and features of genomic organization such as recombination rate and the locations of coding and conserved basepairs, which appear to have consistent effects on ancestry in many species (e.g. *Principles 2 & 3*). Moving beyond these broad scale features, there are good reasons to expect that replicated hybridization events between the same species will lead to predictable outcomes at the genomic level. In repeated hybridization events between the same species, the same genetic interactions and selective forces are predicted to drive concordant changes in ancestry along the genome. Indeed, this has been observed in experimental hybrid populations, natural hybrid populations [28,44,56,75,155–158], and in replicated cline studies [159] (but see [160]).

While it seems sensible to expect that replicated hybridization events should lead to similar patterns of local ancestry, recent work has suggested that in some cases we may expect more repeatability *across* taxa than predicted by classic evolutionary theory [22]. Studies in *Arabidopsis* and *Xiphophorus* have repeatedly uncovered some of the same genes underlying hybrid incompatibilities (Fig. 1; [9,161–163]), and certain genetic interactions, such as cytonuclear incompatibilities, are common across the tree of life [113,164–170]. These results suggest that some types of genetic interactions are more prone to breaking down in hybrids, perhaps due to their function, the rate at which they accumulate substitutions, or their position in a gene network. Whether incompatibilities frequently evolve in the same genes or pathways has important implications for whether we expect regions resistant to introgression to be shared across species.

Similarly, comparisons of dynamics of adaptive introgression across species have identified possible hotspots in terms of gene families that appear to confer advantages when introgressed. This process could also generate fine-scale repeatability in local ancestry. The best documented examples include immune related genes [3,14,171], pigmentation genes [172–175], and genes that underlie resistance [34,149,176,177], but many other functional categories of genes or selfish elements could have similar behavior [178,179].

Compared to incompatibilities and adaptive introgression, we know much less about how other forms of selection on hybrids might lead to predictable outcomes at the local scale. Although it has not been directly studied, selection against hybridization load could lead to partially predictable outcomes across replicated hybridization events. Regions of the genome with lower local $N_e$ (i.e. due to variation in the effects of background selection [180,181]) should accumulate more weakly deleterious mutations within populations and thus be more likely to be purged after hybridization. Additionally, gene dense regions provide a larger target for functionally relevant mutations to occur and may therefore experience stronger selection in the early generations after hybridization when ancestry tracts are long.

For other mechanisms of selection, we expect much lower predictability across systems. For example, if species have independently adapted to distinct ecological conditions, we would not expect the genetic architecture of such traits to be shared except in rare cases (e.g. [182]). Without selection on the same underlying regions of the genome, any repeatability in local ancestry patterns in hybrids should not exceed what is expected due to broad scale features such as gene density (*Principles 2 & 3*).

*Predicting differences in local and global ancestry between species*

Conserved mechanisms that shape ancestry after hybridization can also point to cases where we predict to see differences between species. We recently found differences in the extent to which introgressed haplotypes were retained in coding regions in the genomes of swordtail fish and humans, likely due to differences in the underlying recombination maps [28]. Both species share a strong positive correlation between introgression and the local recombination rate. However, recombination is concentrated in promoters and other functional regions in swordtail fish [87], and tends to occur away from such regions in humans [85,86]. This results in distinct patterns of local ancestry, with swordtail fish retaining more minor parent ancestry than humans in and around genes (presumably due to differing outcomes of the action of *Principles 2 & 3* in the two species groups). As data from more diverse systems accumulates, comparative analyses of patterns of introgression as a function of these features of genome structure, combined with theoretical analyses, will further develop our understanding of how selection acts in admixed genomes.

Similarly, as discussed in *Principle 1*, the speed of initial purging of minor parent ancestry is sensitive to the aggregate recombination rate, which differs widely between species [70]. This is because the aggregate recombination rate is strongly influenced by the total number of chromosomes and whether recombination occurs in both sexes - properties that vary widely across the tree of life [70,183]. Notably, these factors together may be important in explaining the variation in admixture proportions observed in the genomes of different species that are known to commonly hybridize, from cases where retention of minor parent ancestry after hybridization is rare, such as *Drosophila* [184], to those where extensive introgression is common, such as swordtail fish [16].

*Effects of hybrid demography*

As is the case in non-admixed populations, we expect that certain features of genome evolution will be sensitive to the demographic history of hybrid populations themselves. The importance of demography in tension zones has long been appreciated, where dynamics of dispersal and population density at the contact zone [185,186] play a key role in the outcomes of hybridization and interpretation of cline analysis [66,143,187,188]. The impacts of demography on hybridization have been less thoroughly explored in the context of pulses of admixture. However, there are multiple reasons to predict that pulses of hybridization may coincide with strong bottlenecks, since they are often driven by ecological disturbance [189,190] and because selection on hybrids can be so strong that it essentially drives population collapse [191].

Intuitively, the long-term size of hybrid populations and the proportion of parental genetic diversity retained in hybrids should have important impacts on genome evolution. In many cases selection on hybrids will be strong enough to overcome the effects of genetic drift, even in small populations, especially in early generation hybrids when many selected sites are linked. Over long time periods, however, populations with small effective size will be less efficient at purging weakly deleterious variants that occur in short ancestry tracts.

Another important consideration is the number of parental individuals from each species that contributed to a hybridization event, which will shape the raw material on which selection can act. We recently mapped the genetic basis of a hybrid melanoma that develops from a tail pigmentation spot in swordtail fish. Notably, this tail pigmentation spot is polymorphic in one of the parental species (~30% frequency; [162]). Presumably due to differences in the founding parental populations, some hybrid populations have both a high frequency of the tail spot and of melanoma, whereas others have a low frequency of both [162]. Though just one example, this highlights how the genetic contribution of the parental species can be an important element influencing how selection will act within hybrid populations. Studies in other systems such as *Drosophila*, *Mimulus*, and *C. elegans,* have identified polymorphic hybrid incompatibilities, suggesting that these founder dynamics could have an important impacts on hybrid populations [106,192–194].

**Ways forward**

Hybridization often leads to unusually dynamic genome evolution and reorganization, which we are just beginning to understand. As more data become available from diverse hybridization events, across taxa and timescales, we can begin to distinguish between the different processes that shape ancestry in the genome after hybridization. Ultimately, we hope such research will lead to an understanding of how different sources of selection interact with each other and with variables such as genome structure, to drive similarities and differences in patterns of introgression across species. Although there are outstanding questions that may require years to disentangle (see **Box 3**), to conclude our discussion we propose a way forward to tackle a subset of these questions.

*Repeatability in the evolution of hybrid incompatibilities*
In previous sections we discuss the uncertainties surrounding how hybrid incompatibilities arise and the degree to which we expect incompatibilities to arise repeatedly (Fig. 1), either in the same genes [162,163] or in the same regions of the genome [28,31,83]. Such repeatability in the evolution of hybrid incompatibilities could undermine assumptions of the **snowball effect**, which posits that because newly arising mutations in one species can interact with any derived mutations in the second species, the number of incompatibilities between two species should increase exponentially over time [195]. However, if the mutations that cause DMIs are enriched in the same genes or genomic regions, the rate of this accumulation should slow [196,197]. Similar predictions emerge from theoretical studies of gene regulatory network evolution, where the likelihood of a gene's involvement in DMIs is directly related to the density of the gene network [198,199]. Systematic differences in gene network connectivity between species could drive differences in the distribution of DMIs across the tree of life [200]. Though limited by the experimental and statistical challenges inherent in identifying DMIs, both evidence for DMI "hotspots" and a slowed snowball effect should be detectable from empirical data in experiments with sufficient power.

*Distinguishing between selective forces*

The differences in genetic architecture assumed by each model of selection on hybrids is one promising route to inferring their role in shaping local ancestry after hybridization. Selection on DMIs is generally thought to be stronger and less polygenic than hybridization load models (**Box 1**; but see [148] for an exploration of polygenic epistatic selection).  Higher levels of polygenicity will increase the proportion of neutral basepairs that are linked to sites that are deleterious in hybrids (Fig. 4). Moreover, under a DMI model ancestral and transitional genotypes can be favored by selection, which will actually act to increase minor parent ancestry in some regions of the genome (Fig. 4 - Supplement 1). Together these factors will lead to greater purging of minor parent ancestry over time under polygenic models of selection against minor parent ancestry, as opposed to DMI models (Fig. 4). Comparing the predictions of these different architectures of selection on hybrids using modeling or simulations could be a powerful tool to allow researchers to distinguish between them, at least on a genome-wide scale (as in [26]).

*Empirical studies of hybrid evolution*

Studies of selection in contemporary hybridizing populations offer another route to merge pattern and process, and to tease apart forms of selection acting in admixed populations. For example, Chen [201,202] and Fitzpatrick [203] studied weakly differentiated populations and found that genome-wide selection broadly favored ancestry derived from migrants in small populations, consistent with the idea that in small populations foreign ancestry can be favored to lighten the genetic load. In contrast, we recently found little evidence that hybridization load shapes genome-wide ancestry in hybrid swordtail populations formed between species with substantial differences in historical effective population size [28]. While these studies used genomic tracking in natural populations, other researchers have leveraged laboratory crosses and systematically varied environmental conditions to explore how ecological selection shapes genome evolution [204]. Combining such observational and manipulative approaches with comparisons across diverse species may reveal the relative importance of the forces shaping evolution after hybridization along the speciation continuum.

*Predicting differences between species after hybridization*

Examples of hybridization across the tree of life poise the field for a broader analysis of what genetic and biological features are associated with variation in rates of introgression. For one, theory predicts that species with fewer chromosomes will undergo faster and stronger purging of minor parent ancestry in their genomes, due to a low aggregate recombination rate ([70], e.g. in species such as *Arabidopsis*, *Drosophila*, mosquitoes). In addition to empirical analyses to address key theoretical predictions, the wealth of newly available data opens up a large number of possible studies of underexplored features of organismal biology that could influence retention of minor parent ancestry after hybridization, which we discuss briefly here.

Life history traits may play an important role in variation in introgression across the tree of life [205]. For example, the extent of selfing or asexual reproduction impacts the genetic diversity of the parent populations, their genetic load, and the frequency with which recombination reshuffles parental haplotypes, and therefore can shape the extent and direction of introgression [206,207]. Similarly, some data suggest that systems with facultative asexual reproduction can retain larger minor parent contributions [208–210], and tolerance of genome duplication and aneuploidy will interplay with retention or loss of parental genomic material.

Variation in the structure and function of the genome between species may also play a key role. Decades of work have established an important role for inversions in locally restricting gene flow [211–214]. Beyond inversions, the frequency and activity of transposable elements in the genome is a classic mediator of selection against hybrids, but mixed evidence for its generality necessitates broader study [215–217]. Gene expression (or misexpression) that is specific to life cycle stage or tissue type could lead to temporal or tissue-specific fitness effects in hybrids. Notably, recent work has demonstrated that there is weaker selection against Neanderthal ancestry in enhancers that are tissue specific in modern humans [74]. This highlights the potential for such context dependence, which would certainly vary across species groups (e.g. fungi versus plants and animals), and shape how admixed genomes are exposed to the varied forms of selection discussed above. Other features of genome organization that differ widely across species groups such as the presence of micro-chromosomes, polyploidy, and recombination mechanism, will all be rich areas to study in this regard [205,218–224].

**Conclusions**

Though there are major challenges ahead, we have made significant progress in the past decade characterizing the diversity of hybridization events across the tree of life. Here, we hope to have illustrated that our knowledge of the basic processes at play and theoretical predictions about hybrid genome evolution have grown greatly as a product of this work. On a broad scale, genome stabilization after admixture is now known to be a multi-stage process affected by the distribution of functional elements and the recombination landscape. Multiple selective forces may affect genome evolution after hybridization, and the intersection of these forces is ripe for empirical and theoretical investigation. While many outstanding questions remain**,** we are now, more than ever, poised to disentangle the factors impacting genome evolution in hybrids and build new models of how they interact. Research in these areas will lead to a better understanding of the nature of reproductive barriers between species and the genetic and evolutionary impacts of hybridization across the tree of life.

**Glossary**

**Adaptive introgression:** the hybridization-mediated transfer of parental alleles that increase fitness in the recipient population.
**Admixture:** a more general term than hybridization that encompasses all mating between distinct populations, which may or may not be diverged enough to be considered species.
**Ancestry linkage disequilibrium (ancestry LD):** statistical association between haplotypes of the same ancestry, that can be caused by physical linkage of sites, selection, or population structure; in the case of linkage disequilibrium due to physical linkage, ancestry LD extends over much greater physical distances than is typical for non-admixed populations.
**Ecological selection:** Selection driven by the fitness of an organism's traits in the context of its environment.
**Fisher's geometric model:** a general model of selection where fitness is determined by distance from a phenotypic optimum, which has been applied in the hybridization literature to describe selection on polygenic traits (either stabilizing or directional; Fig. 2), ecological selection on hybrids, and hybrid incompatibilities.
**Genetic architecture:** the number, effect size, and location in the genome of loci contributing to a phenotype.
**Haplotype:** a physically contiguous tract of DNA inherited from a single parent unbroken by recombination.
**Hybrid incompatibilities:** mutations which arise in interacting genes after two lineages diverge such that when individuals from these populations hybridize a previously "untested" combination of alleles reduces hybrid viability or fertility.
**Hybridization load:** The burden of mildly deleterious mutations which preferentially accumulated in the parental lineage with less effective selection, leading to reduced fitness of hybrids that harbor more of that species' genome and selection against ancestry derived from that species. Fitness in hybrids is not reduced relative to the parental species with lower historical effective population size.
**Introgression:** Transfer of a region of the genome between species due to hybridization.
**Major parent:** the species that contributed a majority of the genome of an admixed population.
**Minor parent:** the species that contributed a minority of the genome of an admixed population.
**Polygenic trait:** a trait where phenotypic variation is explained by the combined effects of many, sometimes thousands, of variants spread throughout the genome.
**Segregation load:** the decrease in average fitness of hybrids expected due to the disruption of co-adapted sets of alleles inherited from the parental species that are broken apart by recombination and independent assortment.
**Sexual selection:** selection driven by mate choice and competition for mates.
**Snowball effect:** the faster-than-linear increase in the number of DMIs with increasing numbers of substitutions between two species that is predicted by evolutionary theory.

**Species:** Two groups of organisms where hybrids between them have reduced viability or fertility. This can range from moderate impacts on viability or fertility to complete inviability or infertility.

**Hybrid zone:** Spatial zone where hybrids form between the geographic regions occupied by two parental species.

**Tension zone:** A stable zone where hybrids are found in a narrow geographical region, as a result of balance between ongoing dispersal of individuals from parent populations and strong selection against hybrids.

**Box 1 – Predicted outcomes under different sources of selection on hybrids**

Here, we discuss ancestry patterns that are consistent with different modes of selection on hybrids.

***Selection against minor parent ancestry*** – Under the DMI model, loci derived from the minor parent are more likely to uncover incompatibilities elsewhere in the genome, leading to global selection against minor parent ancestry [28]. Similarly, under a model of polygenic selection against hybrids as a function of the disruption of co-adapted parental alleles, loci derived from the minor parent will, on average, result in hybrids whose genotype combinations are further from phenotypic optima. This may result in a genome-wide shift towards major parent ancestry.

***Selection is context dependent*** – In the case of hybridization load, selection is expected to act against ancestry derived from the parental species with lower historical effective population size, whether that is the major or minor parent [26–28]. Likewise, in the case of ecological selection, expected patterns are driven by the ecological environment. If hybrids occur in a habitat most similar to that of the minor parent, selection is expected to favor ancestry from the minor parent, and if hybrids occur in a habitat most similar to that of major parent, selection is expected to favor ancestry from the major parent.

***Unique signals*** – Unlike other models, hybridization load is explicitly limited to weak selection: selection coefficients that are much greater than the reciprocal of the historical effective population size of the parental species are not consistent with the predictions of this model [26,27]. Ecological selection is dependent on the environment, and thus changing the relevant environmental parameters should change the direction of selection [233]. Though technically challenging, empirical studies evaluating the phenotypes of surviving hybrids compared to parentals could predict the traits and ancestry selected by specific environmental conditions.

***Genetic architecture*** – Models of hybridization load and polygenic selection on hybrids tend to envision a scenario in which numerous loci are under weak selection, while DMIs are generally assumed to be stronger and less polygenic. While the validity of some of these assumptions awaits more empirical data, these models should generate distinct predictions about the extent and patterns of purging of minor parent ancestry after hybridization, which have yet to be rigorously characterized (see Fig. 4; **Ways Forward**).

**Box 2 – Complexity introduced by transgressive traits, recessive load, and sexual selection on hybrids**

*Ecological selection and transgressive traits* – While hybrids often have phenotypes that fall within the parental ranges, transgressive traits, or those outside of the distribution observed in either parental species, are also common (approximately 20% of traits in $F_{1s}$ in some studies [123]). Though we might generally expect such traits to be selected against (Fig. 2), transgressive phenotypes are sometimes better suited to novel environments than parental phenotypes, and as a result can promote ecological speciation [19,130,234–239]. Because the genetic divergence between species appears to predict the frequency of transgressive traits, we may also expect to see variation in the frequency of hybrid speciation as a function of parental divergence [240,241]. However, this is complicated by the fact that mechanisms driving selection against hybrids, such as hybrid incompatibilities, are also expected to scale with divergence.

*Recessive Load favoring Introgression* – If deleterious mutations segregating in populations are largely recessive, selection could broadly favor foreign ancestry in admixture between species with similar historical effective population sizes. This is because each diverged population accumulates its own private set of deleterious variants, which will be reciprocally masked by heterozygous ancestry tracts [27,120]. These heterosis dynamics can even mimic the signal of adaptive introgression [121,242].

*Sexual selection* – Often overlooked as a force impacting genome evolution in hybrids, sexual selection acts on hybrids in complex ways that depend on the frequency of both preference and mate choice trait loci in the population [243]. Furthermore, mating preferences are often multivariate [244–247], and recombination can break up trait correlations as well as multimodal preferences [187,245,248], resulting in a variable landscape of sexually selected traits and preferences. The impacts of these recombinant trait and preference phenotypes on ancestry will be largely dependent on the strength and nature of selection exerted by both parental and hybrid females (or males in systems with sex-role reversal), and whether preferences are fundamentally different in hybrid populations.

**Box 3 – Outstanding Questions**

The near-term goals discussed in **Ways Forward** present tractable problems toward which preliminary efforts can be or have been made. Here, we highlight more open-ended questions which will likely take years of further study to address.

**How do we reconcile the evidence of frequent historical gene flow across the tree of life with the evidence for reduced hybrid fitness?** Widespread evidence of historical hybridization in the genomes of modern species suggests that despite strong selection on hybrids (on top of strong prezygotic barriers in many systems), hybrids do persist and reproduce. While this apparent conundrum could be explained in part by the rapid purging of regions of the genome that are deleterious in hybrids, the overall observation remains puzzling, as does the fact that premating barriers which prevent maladaptive hybridization are fragile in so many systems (e.g. [189,249]).

**Are there additional undiscovered variables which contribute to tolerance of introgression?** It has been recently shown that aggregate recombination rate is a key variable impacting permeability of a genome to introgression, providing a novel explanation for the observation that some species have extremely low rates of introgression despite frequent hybridization in nature, including classic models such as *Drosophila* [7]. The observation that fitness of hybrids between pairs of species of a given genetic divergence varies widely across study systems suggests the presence of other, as of yet unknown factors, affecting the strength of selection against hybrids. Whether those factors are the true architecture of selection, the nature of genetic networks, or systematic differences between species (i.e. such as in recombination mechanisms, reproductive system) remains to be seen.

**Which theoretical model(s) best represent selection on hybrids?** Established models of selection provide tractable predictions about introgression patterns but may poorly describe the complexity of biological systems. For example, selection against gene misexpression in hybrids may reflect aspects of both DMIs and stabilizing selection on gene expression. These predictions become even more complicated with conflicting sources of selection acting on hybrids (e.g. Fig. 3) and disentangling them may not always be tractable.

# Figures

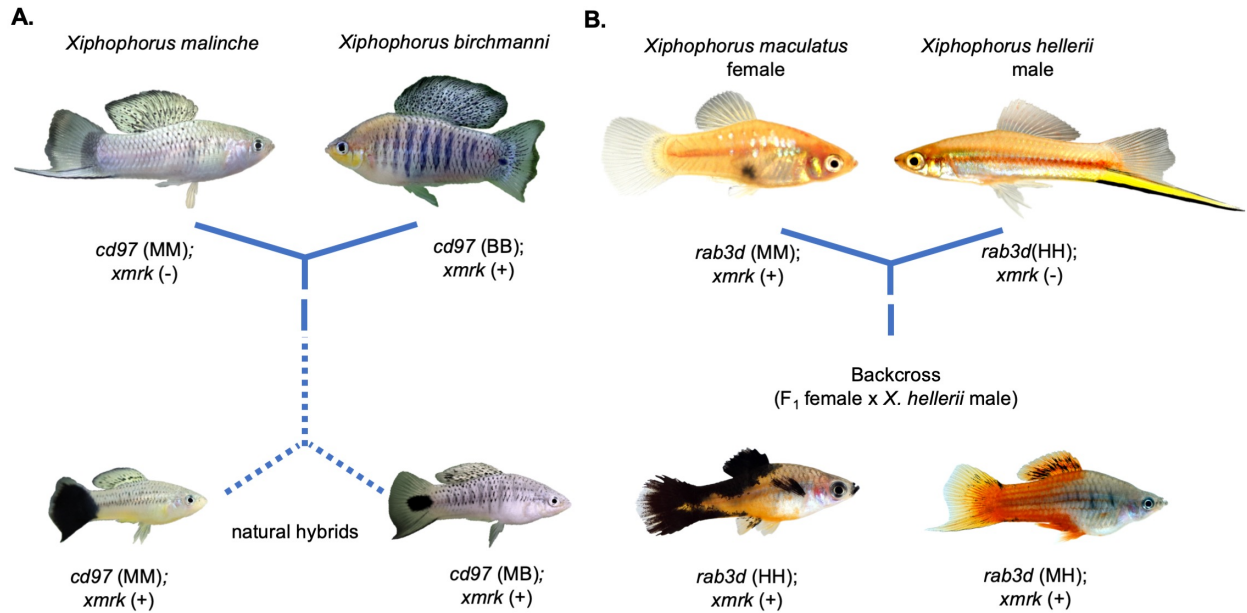

**Fig. 1: Repeated hybrid incompatibilities in *Xiphophorus*.** Classic models in evolutionary biology predict that incompatibilities can arise between any pair of interacting genes. Recent empirical work has suggested that certain genes or pathways may be especially prone to becoming involved in hybrid incompatibilities. The gene *xmrk* independently causes melanoma in hybrids between different swordtail fish species. **A**. In crosses between *X. birchmanni* and *X. malinche*, *xmrk* interacts with the gene *cd97* to generate melanoma in a subset of hybrids [162]. **B**. In crosses between distantly related species *X. maculatus* and *X. hellerii*, *xmrk* interacts with a different region, *rab3d,* to cause melanoma [163,250,251]. Phylogenetic analyses suggest that these incompatibilities with *xmrk* have arisen independently [162]. Photos of hybrids in **B** were provided by Manfred Schartl.

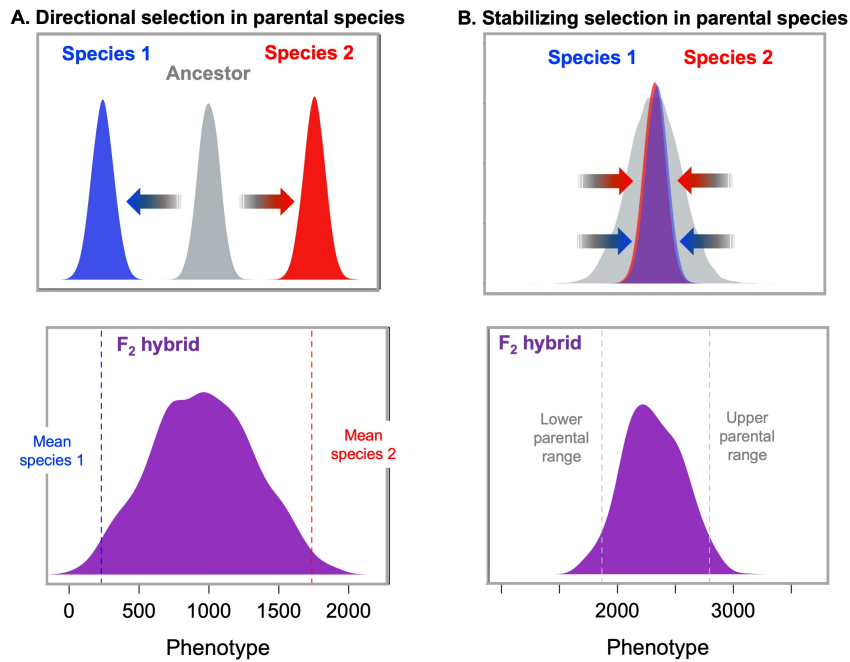

**Fig. 2: Models of selection on polygenic traits in parental species and their implications for hybrids. A. Top -** If two species have adapted from the ancestral state (gray) towards two different phenotypic optima (blue and red respectively), hybrids between those two species (purple, **bottom**) are predicted to fall far from the phenotypic optimum of either parental species [143,252,253]. The distribution shown for $F_2$ hybrids was generated by simulating a phenotype controlled by 10 loci in each of the parental species with an exponential distribution of effect sizes, a mean trait value of 250 for parent species one (dashed blue line), a mean trait value of 1750 for parent species two (dashed red line), and additive effects at each locus on the phenotype. Simulations were performed in admix'em [254]. **B. Top -** Similar principles apply in the case of a polygenic trait that does not differ between the parental species because it has been under stabilizing selection [142,143]. In this case, different combinations of trait increasing and trait decreasing alleles are expected to have fixed over time in the two parental species without changing the average trait value across species. As a result, this will generate increased phenotypic variance in $F_2$ and later generation hybrids compared to the parental species [142,143]. These higher variance phenotypes in hybrids should be selected against via stabilizing selection. Simulations shown here illustrate this principle; $F_2$ hybrids (purple **bottom**) have increased trait variance relative to the parental species. Simulations were performed as above but the average trait value was the same in the two parental species (2200). The underlying alleles and their effect sizes for this simulation were drawn from a random exponential distribution.

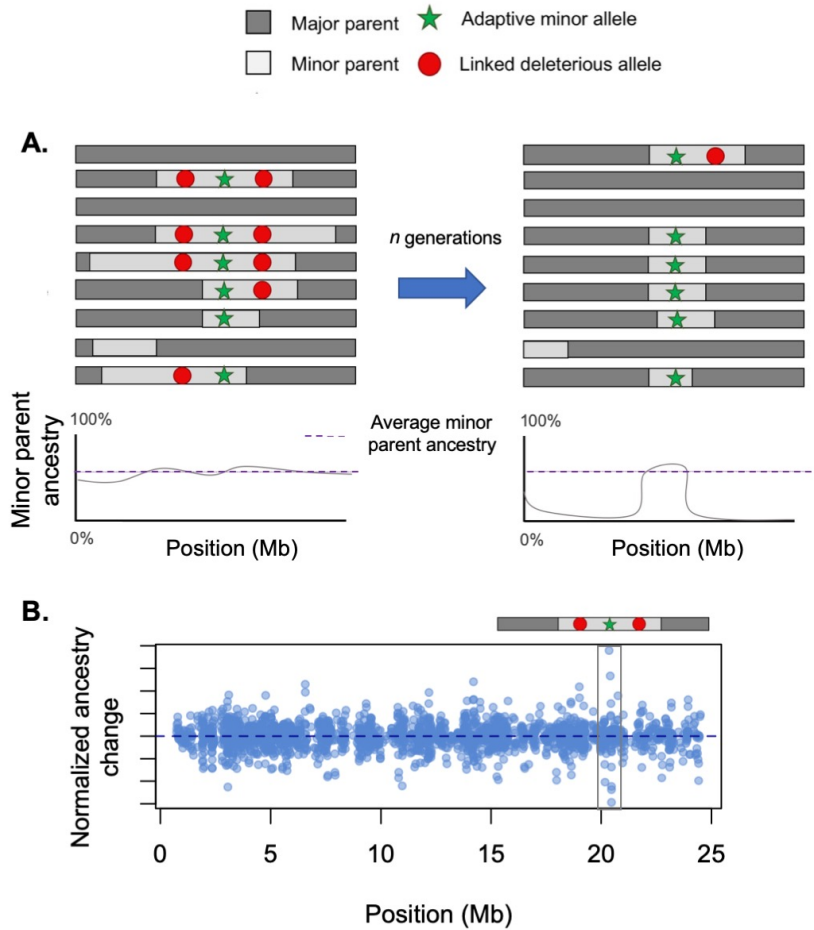

**Fig. 3: Conflicting selection between linked alleles.** Hybridization derived-haplotypes can be deleterious, neutral, or adaptive. In some cases, selection on deleterious and adaptive sites may interfere with each other. **A.** Here, we illustrate a case in which there is tight physical linkage between sites that are deleterious in hybrids (such as DMIs) and a site that is beneficial. **Left –** When positively and negatively selected sites are linked on the same haplotype, selection will act on the average of their selection coefficients. In this case, due to interference between positive and negative selection, ancestry is relatively stable in this region when selected sites are linked to the same haplotype. **Right -** After a recombination event occurs and breaks apart this linkage, the selected haplotype will begin to rapidly increase in frequency. **B.** Although not easily detectable using existing methods, such interference effects are potentially detectable using sharp transitions in ancestry over a short distance. Here we illustrate the results of a simulation using the hybrid population simulator admix'em [254] where an adaptive locus ($s=0.05$) is flanked on either side with loci deleterious in hybrids (each $s= -0.05$, 50 kb away). The admixture proportions simulated were 75% parent 1 and 25% parent 2 and the simulation was conducted for 200 generations. In this simulation a haplotype arises where recombination events have unlinked the adaptive and deleterious sites, allowing the haplotype harboring the adaptive allele to begin to sweep to fixation. Long before fixation has occurred, however, the adaptive haplotype (gray box) is detectable due to the sharp ancestry change surrounding it.

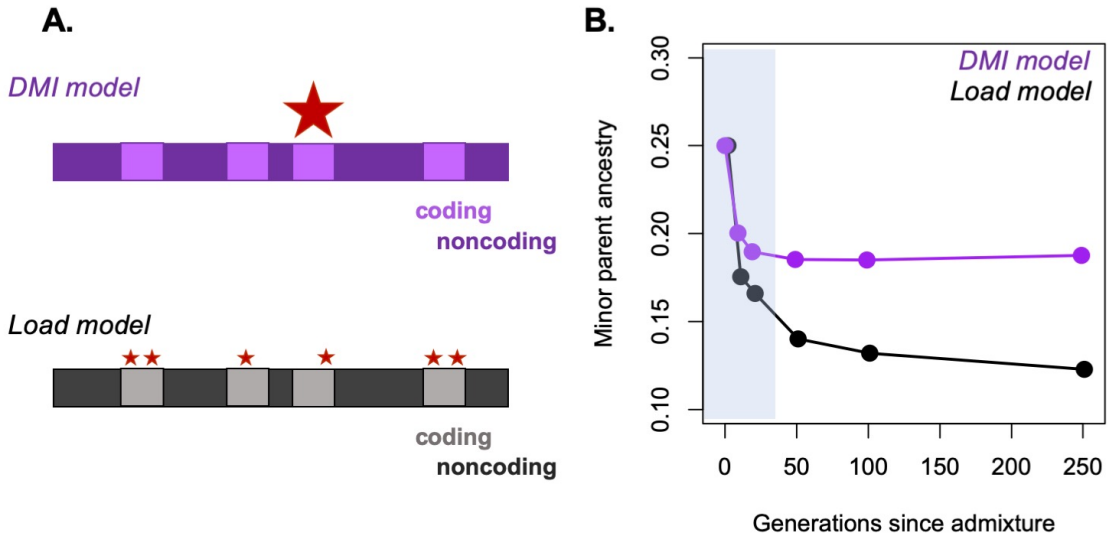

**Fig. 4: Possible approaches to differentiating between selective forces in simulations.** A major challenge in the field is distinguishing between possible sources of selection driving particular patterns of ancestry in hybrids. One promising approach is to use simulations to begin to distinguish between these possibilities. **A.** As an example, we simulate ancestry change under two models, the DMI model and the hybridization load model after a pulse of admixture. Selected sites are shown as red stars, with the size of the star in the schematic corresponding to the strength of selection on individual sites. **B**. We performed simulations using SLiM under these two models of selection on hybrids [255]. Admixture proportions for both simulations were set at 75% parent 1 and 25% parent 2, and $F_1$ fitness in hybrids was 0.85. Ancestry was tracked on a 25 Mb chromosome in a diploid hybrid population (N=2,000). In the simulation shown in purple, selection on hybrids is driven by selection on three hybrid incompatibilities with dominance of 0.5, randomly positioned along the chromosome. In the simulation shown in black, selection on hybrids mimics a load model, with a total of 160 sites under selection along the chromosome. In this case, the selected sites are deleterious in all genetic backgrounds. The shaded area indicates the period of "fast" initial purging (*Principle 1*) which is followed by a slower period of long-term purging in the hybridization load simulation. Although differences in the dynamics of purging between the two models are partly driven by the number of loci under selection in hybrids, the DMI model differs from other models of selection because not all minor parent alleles are disfavored (see Fig. 4 – Supplement 1).

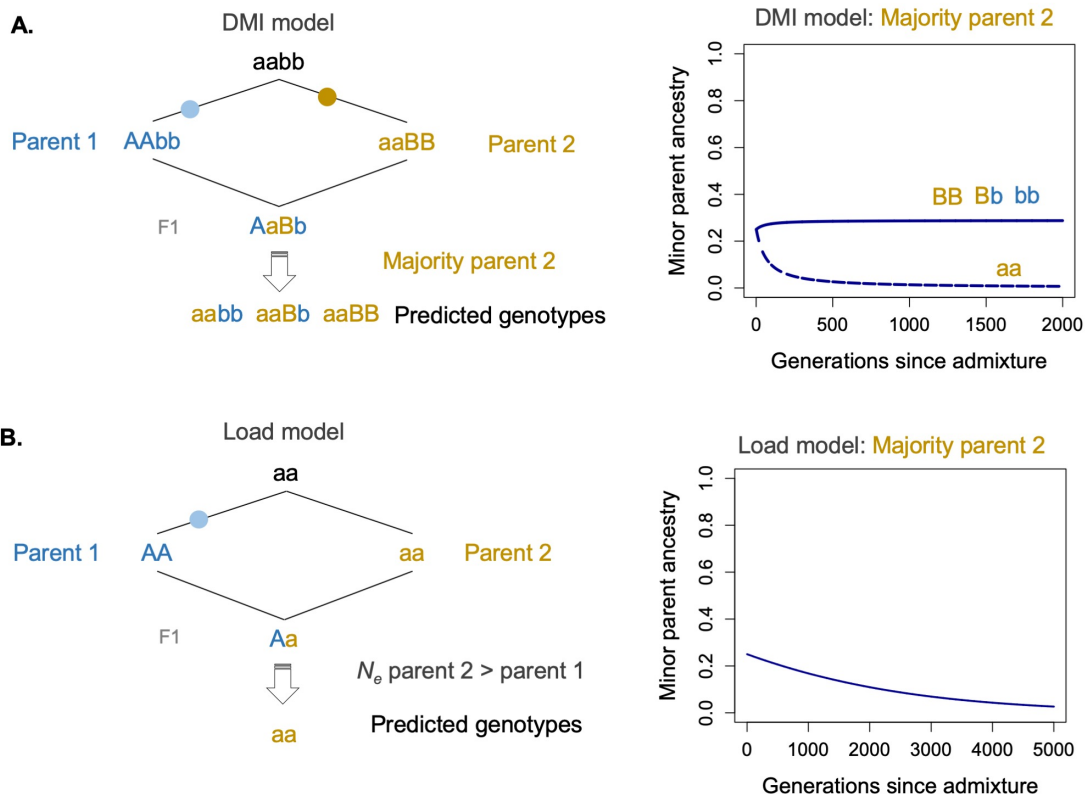

**Fig. 4 – Supplement 1.** The DMI model and model of selection against hybridization load differ not just in assumptions about the number of sites under selection and strength of selection coefficients but also in the types of genotypes under selection. **A.** Specifically, under a two-locus DMI model, minor parent ancestry at one of the loci can be neutral or even favored by selection. This is because recombination between the parental types can regenerate ancestral (aabb) and transitional genotypes (AAbb,aaBB), which are globally compatible (left panel). Models predict that this will lead to retention of minor parent ancestry at one of the two loci involved in the DMI ([109,256], right panel). Solid line shows ancestry at locus 1 and dashed line show ancestry at locus 2 under a deterministic model of selection on hybrid incompatibilities [256] with a starting admixture proportion of 0.25, selection coefficient of 0.1, and dominance of 0.5. **B.** In contrast, when minor parent ancestry is globally deleterious at sites under selection, purging is always expected. An example is shown in the left panel of fixation of a deleterious site in parent species 1 as a result of less efficient selection due to lower historical effective population sizes. After admixture (right panel), parent 1 ancestry is predicted to be purged at this locus, assuming that hybrid population sizes are sufficiently large. See the main text for a detailed discussion of this issue. Solid line shows ancestry at the selected locus with a starting admixture proportion of 0.25, selection coefficient of 0.001, and dominance of 0.5.

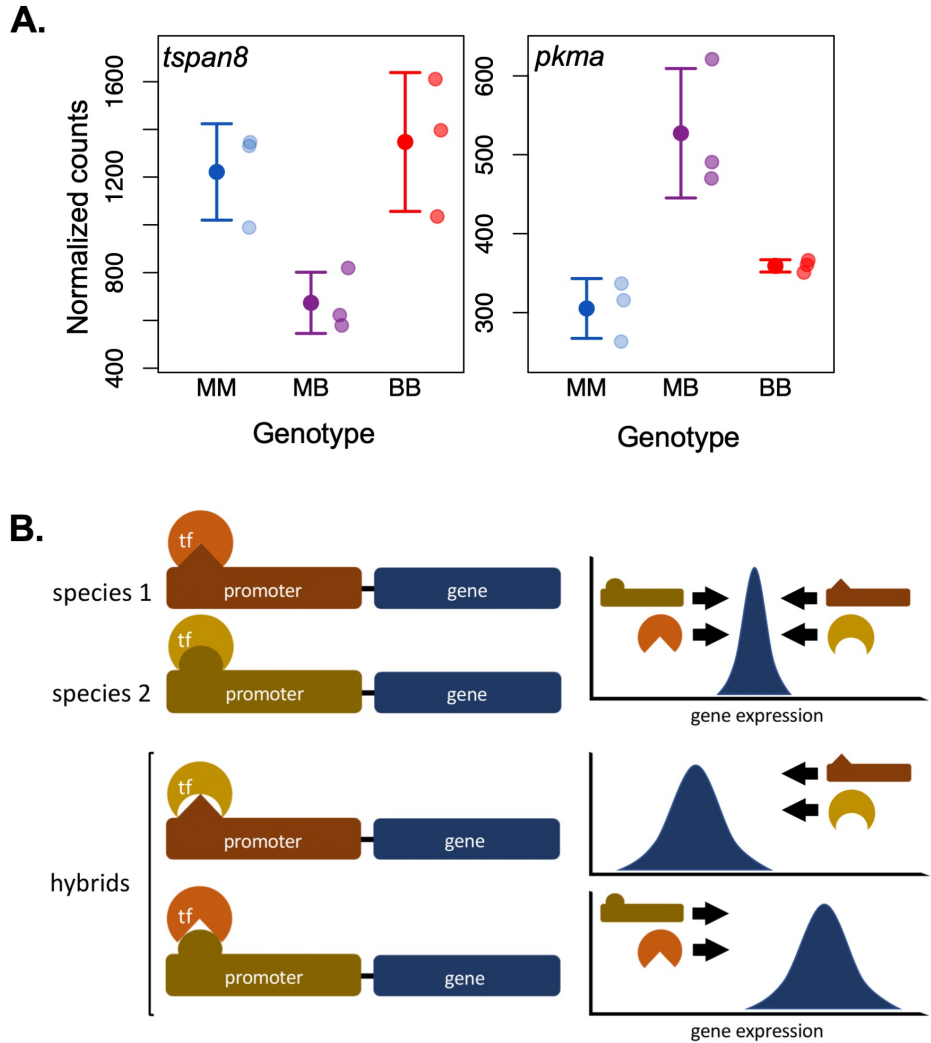

**Fig. 5: Selection on gene expression in hybrids.** Hybridization can generate mismatch between *cis*- and *trans*-acting regulatory factors that have co-evolved within the parental lineages to regulate expression of target genes around an expression optimum (i.e. through stabilizing selection). This can result in an incompatibility generated by misregulation and transgressive expression of such genes in hybrids. **A.** *tspan8* (left) and *pkma* (right) are examples of genes for which swordtail hybrids exhibit low and high misexpression, respectively (MM – *X. malinche*, BB – *X. birchmanni*, MB – $F_1$ hybrids; data from [257]). **B.** This simplified diagram illustrates how mismatches in co-evolved regulatory elements can cause misexpression. Promoters and transcriptions factors (TFs) are a classic example of *cis* and *trans* regulatory elements that interact to promote or suppress expression of target genes. Promoters and TFs can evolve to have opposing regulatory effects on target genes to achieve optimal expression (top), leading to differences in structure, interacting residues, or binding affinity between diverged populations. In hybrids, divergent binding sites within the promoter and changes in binding affinity of the TF may result in over or under expression of target genes, leading to misexpression (bottom).

**Acknowledgements**
We thank Nico Bierne, Erin Calfee, Kelley Harris, Bernard Kim, Erica Larson, Pavitra Muralidhar, Greg Owens, Yuval Simons, Scott Taylor, Ken Thompson, Carl Veller, and members of the Schumer, Brandvain, and Matute labs for helpful discussion or feedback on earlier versions of this work. Stanford University and the Stanford Research Computing Center provided computational support for this project. This work was supported by a Knight-Hennessy Scholars fellowship and NSF GRFP 2019273798 to B. Moran, a CEHG fellowship and NSF PRFB 2010950 to Q. Langdon, NSF DEB grants 1753632 and 1754246 to Y. Brandvain, and a Hanna H. Gray fellowship, Human Frontiers in Science Young Investigator Award, and NIH 1R35GM133774 grant to M. Schumer.